\documentclass[letterpaper]{article} % For LaTeX2e
\pdfoutput=1
\usepackage{iclr2018_workshop,times}
\usepackage[utf8]{inputenc} % allow utf-8 input
\usepackage[T1]{fontenc}    % use 8-bit T1 fonts
\usepackage{varioref}       % Fancy cross-referencing
\usepackage[pdftex, hidelinks,
            pdfauthor={Richard Wei, Lane Schwartz, and Vikram Adve},
            pdftitle={DLVM: A modern compiler infrastructure for deep learning systems},
            pdfsubject={A neural network compiler infrastructure},
            pdfkeywords={deep learning, 
                         neural networks, 
                         domain-specific languages, 
                         reverse-mode automatic differentiation,
                         reverse-mode algorithmic differentiation, 
                         AD, 
                         optimizing compiler,
                         adjoint code generation,
                         source code transformation,
                         LLVM,
                         DLVM}
          ]{hyperref}       % hyperlinks
\usepackage{url}            % simple URL typesetting
\usepackage{booktabs}       % professional-quality tables
\usepackage{amsfonts}       % blackboard math symbols
\usepackage{bm}             % bold math symbols
\usepackage{nicefrac}       % compact symbols for 1/2, etc.
\usepackage{microtype}      % microtypography
\usepackage{textcomp}       % access \textquotesingle
\usepackage{todonotes}      % todo notes
\usepackage{lipsum}         % placeholder text
\usepackage{graphicx}       % graphics
\usepackage{tikz}           % Modern drawing
\usepackage{fancyvrb}       % Fancy verbatim
\usetikzlibrary{calc, shapes, backgrounds, positioning, arrows.meta, fit, backgrounds}

\definecolor{myblue}{RGB}{93,146,187}
\definecolor{mygreen}{RGB}{89,148,100}
\definecolor{myorange}{RGB}{220,135,81}
\definecolor{myred}{RGB}{200,86,94}
\definecolor{mygrey}{RGB}{118,128,137}

\definecolor{archLightGreen}{RGB}{52,121,116}
\definecolor{archDarkGreen}{RGB}{39,87,87}
\definecolor{archLightBlue}{RGB}{51,116,182}
\definecolor{archDarkBlue}{RGB}{30,76,126}
\definecolor{archBurgundy}{RGB}{91,24,68}

\definecolor{xcodeComment}{RGB}{11,127,10}
\definecolor{xcodeKeyword}{RGB}{25,51,187}
\definecolor{xcodeType}{RGB}{68,147,154}

\title{DLVM: A modern compiler infrastructure for \\ deep learning systems}
\author{Richard Wei \\
  Departments of Computer Science \& Linguistics\\
  University of Illinois at Urbana-Champaign\\
  Urbana, IL 61801 \\
  \texttt{xwei12@illinois.edu} \\
  \And
  Lane Schwartz \\
  Department of Linguistics \\
  University of Illinois at Urbana-Champaign\\
  Urbana, IL 61801 \\
  \texttt{lanes@illinois.edu} \\
  \And
  Vikram Adve \\
  Department of Computer Science\\
  University of Illinois at Urbana-Champaign\\
  Urbana, IL 61801 \\
  \texttt{vadve@illinois.edu} \\
}

\begin{document}

\maketitle

% TLDR:
%
% We introduce a novel compiler framework with a specialized IR that addresses shortcomings of existing deep learning frameworks.

\begin{abstract}
%%% OLD ABSTRACT %%%
% Many current approaches to deep learning make use of high-level frameworks such as TensorFlow, Torch, or Caffe.
% %
% Frameworks such as Caffe have a layer-based programming framework with hard-coded gradients specified for each layer type, making research using novel layer types problematic.
% %
% Frameworks such as Torch and TensorFlow define a computation graph in a host language such as Python, where each node represents a linear algebra operation parallelized as a compute kernel on GPU and stores the result of evaluation;
% %
% some of these frameworks subsequently perform runtime interpretation over that graph, storing the results of forward calculations and reverse-accumulated gradients at each node.
% %
% This approach is more flexible, but these frameworks take a very limited and ad-hoc approach to performing optimization.
% %
% Also problematic are the facts that most frameworks lack type safety, and target only a single (usually GPU) architecture, limiting users' abilities to make use of heterogeneous and emerging hardware architectures.
% %
% We introduce a novel framework for high-level programming that addresses all of the above shortcomings.

Deep learning software demands reliability and performance.
However, many of the existing deep learning frameworks are software libraries that act as an unsafe DSL in Python and a computation graph interpreter.
We present DLVM, a design and implementation of a compiler infrastructure with a linear algebra intermediate representation, algorithmic differentiation by adjoint code generation, domain-specific optimizations and a code generator targeting GPU via LLVM.
Designed as a modern compiler infrastructure inspired by LLVM, DLVM is more modular and more generic than existing deep learning compiler frameworks, and supports tensor DSLs with high expressivity.
With our prototypical staged DSL embedded in Swift, we argue that the DLVM system enables a form of modular, safe and performant frameworks for deep learning. 
\end{abstract}

\section{Introduction \label{sec:intro}}

Within the deep learning community, most current approaches to neural networks make use of high-level frameworks with a tensor domain-specific language (DSL) such as Torch \citep{torch}, TensorFlow \citep{tensorflow}, PyTorch \citep{pytorch}, and MXNet \citep{mxnet}. Traditionally, developers would build a computation graph (or dynamically generate graph nodes) using a DSL and let the framework interpret the computation graph on parallel architectures such as NVIDIA GPUs. While using hand-tuned GPU subroutines usually yields the best performance for complex operators, advanced compiler techniques can be applied to simplify computation, merge high-level operators based on shaping conditions, and fuse compatible element-wise operators to a single kernel to minimize the latency between kernel launches. Recent projects, the TensorFlow XLA compiler \citep{xla} and the NNVM compiler \citep{nnvm} including TVM \citep{tvm}, have begun to apply compiler techniques to deep learning systems, targeting LLVM \citep{llvm} and various back-ends to achieve good performance. However, their design and implementation have not entirely followed established best practices in widely-used compiler frameworks in the industry.

Moreover, some frameworks use operator-overloading algorithmic differentiation (AD) to compute gradients, leaving the gradient computation unoptimizable. The other approach to AD, source code transformation, can produce more efficient code. While frameworks such as TensorFlow already perform AD as a graph transformation and apply various optimizations, their AD transformation is not designed as a transformation pass in the pipeline of their compiler framework, but as part of the DSL library. Making AD part of the compiler framework would greatly simplify the development of DSLs, achieving separation of concerns.

%%%%%%%%%%%%%%%%%%%%%%%%%%%%%%%%%%%%%%%%%%%%%%%%%%%%%%%%%%%%%%%%%%%%%%%%%%%%%%%%
%                                                                              %
%                         Figure: DLVM software stack                          %
%                                                                              %
                              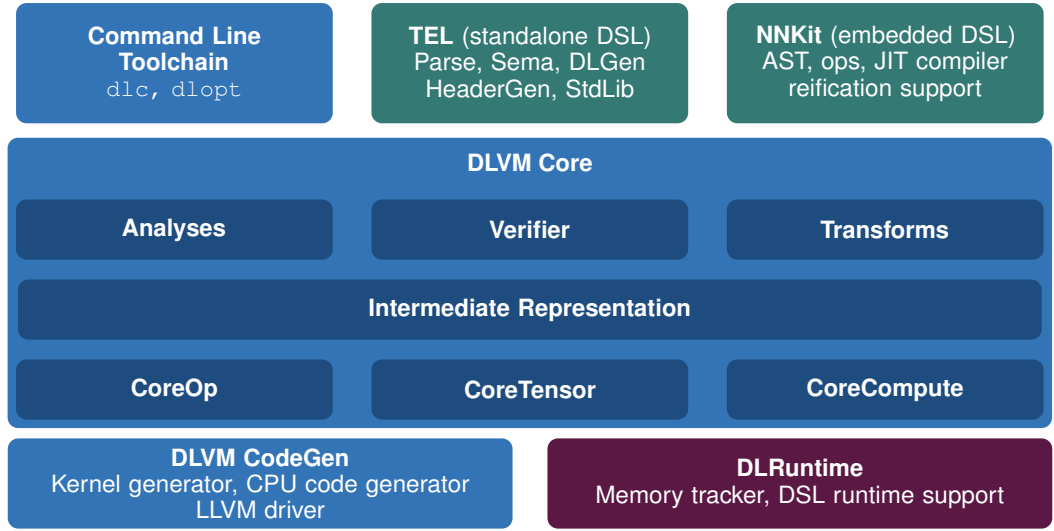
\begin{figure}
\begin{tikzpicture}
\footnotesize
\sffamily
\node (CommandLine)        [               fill = archLightBlue, text=white, shape = rectangle, rounded corners, align=center, text width = 40mm, minimum height=16mm] {\textsf{\textbf{Command Line} \\ \textbf{Toolchain} \\ \texttt{dlc, dlopt}}} ;

\node (TEL)        [ right=5mm of CommandLine, fill = archLightGreen, text=white, shape = rectangle, rounded corners, align=center, text width = 40mm, minimum height=16mm] {\textbf{TEL} (standalone DSL) \\ {Parse, Sema, DLGen} \\ {HeaderGen, StdLib}} ;

% \node (betweenTELNNKit) [ right=2.5mm of TEL] {};

% \node (aboveTELNNKit) [ above=10mm of betweenTELNNKit, text=white] {\textbf{DSL stack}};

\node (NNKit)        [ right=5mm of TEL, fill = archLightGreen, text=white, shape = rectangle, rounded corners, align=center, text width = 40mm, minimum height=16mm] {\textbf{NNKit} (embedded DSL) \\ {AST, ops, JIT compiler} \\ {reification support}} ;

% \begin{scope}[on background layer]
% \node (DSL)        [ fit={(TEL) (NNKit) (aboveTELNNKit)}, fill = archDarkGreen, text=white, shape = rectangle, rounded corners, minimum height=40] {} ;
% \end{scope}

\node (Analyses)        [ below=10mm of CommandLine, fill = archDarkBlue, text=white, shape = rectangle, rounded corners, align=center, text width = 40mm, minimum height=8mm] {\textbf{Analyses}} ;
\node (Verifier)        [ below=10mm of TEL,         fill = archDarkBlue, text=white, shape = rectangle, rounded corners, align=center, text width = 40mm, minimum height=8mm] {\textbf{Verifier}} ;
\node (Transforms)      [ below=10mm of NNKit,       fill = archDarkBlue, text=white, shape = rectangle, rounded corners, align=center, text width = 40mm, minimum height=8mm] {\textbf{Transforms}} ;

\node (Intermediate)    [ below=2.5mm of Verifier,    fill = archDarkBlue, text=white, shape = rectangle, rounded corners, align=center, text width = 134mm, minimum height=8mm] {\textbf{Intermediate Representation}} ;

\node (CoreTensor)        [ below=2.5mm of Intermediate,         fill = archDarkBlue, text=white, shape = rectangle, rounded corners, align=center, text width = 40mm, minimum height=8mm] {\textbf{CoreTensor}} ;
\node (CoreOp)        [ left=5mm of CoreTensor, fill = archDarkBlue, text=white, shape = rectangle, rounded corners, align=center, text width = 40mm, minimum height=8mm] {\textbf{CoreOp}} ;
\node (CoreCompute)     [ right=5mm of CoreTensor,       fill = archDarkBlue, text=white, shape = rectangle, rounded corners, align=center, text width = 40mm, minimum height=8mm] {\textbf{CoreCompute}} ;

\node (aboveVerifier) [ above=2.5mm of Verifier, text=white] {\textbf{DLVM Core}};

\begin{scope}[on background layer]
\node (DLVMCore)        [ fit={(Analyses) (Verifier) (Transforms) (Intermediate) (CoreTensor) (CoreOp) (CoreCompute) (aboveVerifier)}, fill = archLightBlue, text=white, shape = rectangle, rounded corners, minimum height=40] {} ;
\end{scope}

\node (aboveCoreOp) [ above=0mm of CoreOp ] {};

\node (rightOFCoreOp) [ right=9.15mm of aboveCoreOp ] {};

\node (CodeGen)     [ below=10.5mm of rightOFCoreOp,       fill = archLightBlue, text=white, shape = rectangle, rounded corners, align=center, text width = 65mm, minimum height=12mm] {\textbf{DLVM CodeGen} \\ {Kernel generator, CPU code generator} \\ {LLVM driver}} ;

\node (Runtime)     [ right=4.5mm of CodeGen,       fill = archBurgundy, text=white, shape = rectangle, rounded corners, align=center, text width = 65mm, minimum height=12mm] {\textbf{DLRuntime} \\ {Memory tracker, DSL runtime support}} ;

\end{tikzpicture}
\caption{Software stack of the DLVM infrastructure. Blue components are the compiler framework. \label{fig:dlvmStack}}
\end{figure}

%                                                                              %
%%%%%%%%%%%%%%%%%%%%%%%%%%%%%%%%%%%%%%%%%%%%%%%%%%%%%%%%%%%%%%%%%%%%%%%%%%%%%%%%

We introduce DLVM, a new compiler infrastructure for deep learning systems that addresses shortcomings of existing deep learning frameworks.
Our solution includes 
(1) a domain-specific intermediate representation specifically designed for tensor computation,
(2) principled use of modern compiler optimization techniques to substantially simplify neural network computation, including algebra simplification, AD checkpointing, compute kernel fusion, and various traditional compiler optimizations,
(3) code generation through a mature compiler infrastructure that allows for transparent targeting of various hardware, and
(4) an embedded DSL that supports static analysis, type safety, and natural expression of tensor computation, and has a just-in-time (JIT) compiler targeting DLVM for AD, optimizations, and code generation.

\section{Related Work \label{sec:related}}

Numerous existing projects provide specialized systems for machine learning but are not closely related to our work.
These include Apache SystemML \citep{SystemML}, a high-level language and framework for writing and executing machine learning problems targeting Apache Spark, and TACO \citep{TACO}, a C++ library for compiling and optimizing kernels that is more similar to Halide \citep{halide} than to our work. 
Our work treats the creation of neural networks as a compilers problem to be addressed using mature compiler techniques. 
SystemML does not consider this issue at all; TACO does use compiler optimization, but only at a very low level to generate individual kernels.

% SystemML does not consider this issue at all. TACO does use compiler optimization, but only at a very low level to generate individual kernels. There is existing work that is related to ours, namely XLA, TVM, and NNVM. In Section 2, we examine those systems and describe in detail how our work differs from those.

Two projects closely related to this work are the TensorFlow XLA compiler and the NNVM compiler. 
The code representation in these frameworks is a ``sea of nodes'' representation, embedding control flow nodes and composite nodes in a data flow graph. To apply algorithmic differentiation on this IR requires non-standard processing.
In contrast, our approach is designed from the start around the idea that a neural network (and its associated tensor computations) is itself a program, which is best optimized through robust application of mature techniques in a principled compilation pipeline.
We represent tensor computation in static single assignment (SSA) form with control flow graph, and perform algorithmic differentiation, domain-specific optimizations, general-purpose optimizations, low-level optimizations, and code generation.

XLA takes a similar approach to ours, transforming TensorFlow sub-graphs to XLA's HLO graph and performing optimizations.
Our intermediate representation is much more expressive than XLA's by including modular IR components and general-purpose instructions; this enables our approach to support full-fledged DSLs including standalone compiled DSLs and perform more extensive optimizations such as inlining and interprocedual optimizations.
Our approach also differs from XLA by representing composite functions such as \texttt{min} and \texttt{max} directly through primitive instructions such as \texttt{compare} and \texttt{select}, which enables us to apply generic AD, and by using SSA form with control flow graph, which allows for reusing battle-tested SSA optimization algorithms in the LLVM community.
Importantly, our entire infrastructure was designed from the start around a robust compile-time framework for tensor DSLs, whereas XLA has been adapted around the existing TensorFlow infrastructure with a particular focus on hardware support for Google's Tensor Processing Units \citep{TPU}. 

Where TVM and NNVM are built as a DSL and a graph library in Python with a C++ implementation, DLVM's architecture is closer to LLVM and the Swift Intermediate Language \citep{sil}, having an IR file format and a full-fledged command line toolchain.
More specifically, our work differs from NNVM and XLA in the design and presence of an IR that has a textual parsable format, a module/function/basic block hierarchy, custom type declarations and memory semantics. The textual IR enables robust unit testing via FileCheck, which is used extensively in LLVM and most LLVM-based compilers.
Moreover, DLVM and its associated DSLs are implemented entirely in Swift, a safe systems programming language, and thus have an elegantly compact codebase and type-safe APIs.
%

%%%%%%%%%%%%%%%%%%%%%%%%%%%%%%%%%%%%%%%%%%%%%%%%%%%%%%%%%%%%%%%%%%%%%%%%%%%%%%%%
%                                                                              %
%                       Figure: Stages of DLVM pipeline                        %
%                                                                              %
                             \begin{figure}
\begin{tikzpicture}
\footnotesize
\sffamily
\node (analyses)        [                              fill = myblue, text=white, shape = rectangle, rounded corners, align=center, text width = 17mm, minimum height=12mm] {Analyses \& Verification} ;
\node (analyses1)       [below=2mm of analyses,        fill = myblue, text=white, shape = rectangle, rounded corners, align=center, text width = 17mm, minimum height=4mm]  {\scriptsize Dominance} ;
\node (analyses2)       [below=1.4mm of analyses1,       fill = myblue, text=white, shape = rectangle, rounded corners, align=center, text width = 17mm, minimum height=4mm]  {\scriptsize Side Effects} ;
\node (analyses3)       [below=1.4mm of analyses2,       fill = myblue, text=white, shape = rectangle, rounded corners, align=center, text width = 17mm, minimum height=4mm]  {\scriptsize Type Checking} ;
\node (analyses4)       [below=1.4mm of analyses3,       fill = myblue, text=white, shape = rectangle, rounded corners, align=center, text width = 17mm, minimum height=4mm]  {\scriptsize Differentiability} ;

\node (differentiation) [right=5mm of analyses,        fill = myblue, text=white, shape = rectangle, rounded corners, align=center, text width = 18mm, minimum height=12mm] {Differentiation} ;

\node (optimizations)   [right=5mm of differentiation, fill = mygreen, text=white, shape = rectangle, rounded corners, align=center, text width = 20mm, minimum height=12mm] {Optimization} ;
\node (optimizations1)  [below=2mm of optimizations,   fill = mygreen, text=white, shape = rectangle, rounded corners, align=center, text width = 20mm, minimum height=4mm] {\scriptsize Algebra Simplif.} ;
\node (optimizations2)  [below=1.4mm of optimizations1,  fill = mygreen, text=white, shape = rectangle, rounded corners, align=center, text width = 20mm, minimum height=4mm] {\scriptsize AD Checkpointing} ;
\node (optimizations3)  [below=1.4mm of optimizations2,  fill = mygreen, text=white, shape = rectangle, rounded corners, align=center, text width = 20mm, minimum height=4mm] {\scriptsize Linear\,Alg.\,Fusion} ;
\node (optimizations4)  [below=1.4mm of optimizations3,  fill = mygreen, text=white, shape = rectangle, rounded corners, align=center, text width = 20mm, minimum height=4mm] {\scriptsize Dead Code Elim.} ;
\node (optimizations5)  [below=1.4mm of optimizations4,  fill = mygreen, text=white, shape = rectangle, rounded corners, align=center, text width = 20mm, minimum height=4mm] {\scriptsize Comm.\,Subexp.\,Elim.} ;

\node (computegen)      [right=5mm of optimizations,   fill = myorange, text=white, shape = rectangle, rounded corners, align=center, text width = 16mm, minimum height=12mm] {Compute Generation} ;

\node (computsched)     [right=5mm of computegen,      fill = myred, text=white, shape = rectangle, rounded corners, align=center, text width = 16mm, minimum height=12mm] {Compute Scheduling} ;

\node (llgen)           [right=5mm of computsched,     fill = mygrey, text=white, shape = rectangle, rounded corners, align=center, text width = 16mm, minimum height=12mm] {Code Generation} ;

%\node [right=1mm of analyses, single arrow,thick,fill=gray!30] {};

\draw[-{Latex[length=2mm,width=3mm]}, line width=1mm,draw=gray!40] (analyses) -- (differentiation);
\draw[-{Latex[length=2mm,width=3mm]}, line width=1mm,draw=gray!40] (differentiation) -- (optimizations);
\draw[-{Latex[length=2mm,width=3mm]}, line width=1mm,draw=gray!40] (optimizations) -- (computegen);
\draw[-{Latex[length=2mm,width=3mm]}, line width=1mm,draw=gray!40] (computegen) -- (computsched);
\draw[-{Latex[length=2mm,width=3mm]}, line width=1mm,draw=gray!40] (computsched) -- (llgen);

%\node (raw) [above=2mm of analyses, text=black] {\underline{Raw stage}} ;
%\node (opt) [above=2mm of optimizations, text=black] {\underline{Optimizable stage}} ;
%\node (com) [above=2mm of computegen, text=black] {\underline{Compute stage}} ;
%\node (sch) [above=2mm of computsched, text=black] {\underline{Schedule stage}} ;

%\node (opt) [right=17mm of raw, text=black] {\underline{Optimizable stage}} ;
%\node (com) [right=26mm of opt, text=black] {\underline{Compute stage}} ;
%\node (sch) [right=5mm of com, text=black] {\underline{Schedule stage}} ;

\end{tikzpicture}
\caption{Compilation stages in the DLVM compilation pipeline. \label{fig:stages}}
\end{figure}
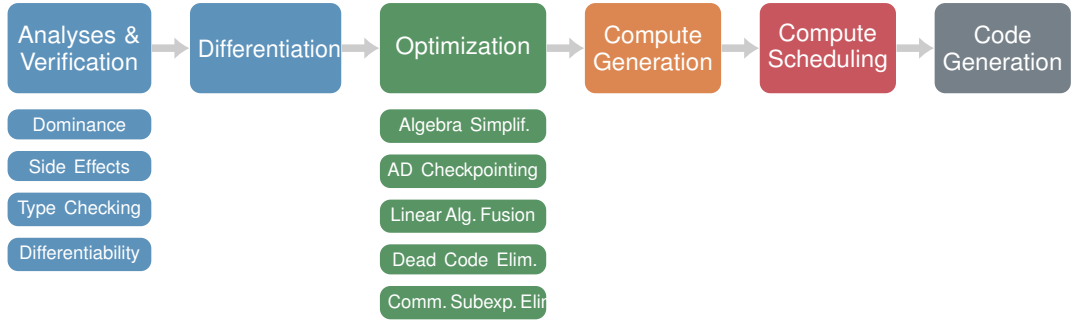

%                                                                              %
%%%%%%%%%%%%%%%%%%%%%%%%%%%%%%%%%%%%%%%%%%%%%%%%%%%%%%%%%%%%%%%%%%%%%%%%%%%%%%%%
\section{DLVM \label{sec:dlvm}}

Deep Learning Virtual Machine (DLVM) is a compiler infrastructure designed for modern deep learning systems.%
\footnote{Code for DLVM is available at \url{https://github.com/dlvm-team}}
DLVM is designed to apply a multi-stage compiler optimization strategy to both high-level linear algebra and low-level parallelism, perform domain-specific transformations, relieve the overhead in front-end languages, and serve as the host for research and development of DSLs for deep learning.
The complete DLVM software stack, including sample front-end deep learning DSLs, is shown in Figure~\vref{fig:dlvmStack}.

Figure~\vref{fig:stages} illustrates the major stages in the DLVM compilation pipeline.
The DLVM compilation stages address algorithmic differentiation, domain-specific optimizations, general-purpose optimizations, and static code generation targeting a variety of compute architectures.
The \textit{raw} and \textit{optimizable} stages allow constructs for high-level tensor operations and various high-level optimizations.
The \textit{compute} and \textit{schedule} stages allow constructs for low-level array operations lowered from tensor operations in high-level stages, borrowing the design from Halide \citep{halide}.

The DLVM Intermediate Representation (IR) is the core language of the system.
It uses static single assignment (SSA) form, control flow graphs, high-level types including a first-class tensor type, and a set of linear algebra operators combined with a general-purpose instruction set (see Table~\ref{tab:dlvmInstructionSet}).
The system enables a wide variety of domain-specific analyses and
transformations, such as reverse-mode AD, AD checkpointing, algebra simplification and linear algebra fusion.

%DLVM IR is separated to four stages: raw, optimizable, compute and schedule. 

%

To demonstrate how DLVM helps the development of domain-specific languages (DSLs), in Section~\ref{sec:DSLs} we present one prototypical staged DSL: NNKit.
NNKit features safe and natural expression of tensor computation alongside the host program, and targets DLVM for differentiation, optimizations and static code generation.

%\todo[inline,color=red!30]{Richard: Revise this section.}

%%%%%%%%%%%%%%%%%%%%%%%%%%%%%%%%%%%%%%%%%%%%%%%%%%%%%%%%%%%%%%%%%%%%%%%%%%%%%%%%
%                                                                              %
%                         Table: DLVM instruction set                          %
%                                                                              %
                          \begin{table}
\begin{center}
\begin{tabular}{|l|p{103mm}|}
\hline
\textbf{Kind} & \textbf{Example} \\
\hline
Element-wise unary   & \texttt{tanh \%a:\ {\textless}{10} x {f32}{\textgreater}} \\
Element-wise binary  & \texttt{power \%a:\ {\textless}{10} x {f32}{\textgreater}, \%b:\ {2}:\ {f32}} \\
Dot                  & \texttt{dot \%a:\ {\textless}{10} x {20} x {f32}{\textgreater}, \%b:\ {\textless}{20} x {2} x {f32}{\textgreater}} \\
%Concatenate          & \texttt{concat \%a:\ {\textless}{10} x {f32}{\textgreater}, \%b:\ {\textless}{20} x {f32}{\textgreater} along 0} \\
Reduce               & \texttt{reduce \%a:\ {\textless}{10} x {30} x {f32}{\textgreater} by add along 1} \\
Transpose            & \texttt{transpose \%m:\ {\textless}{2} x {3} x {4} x {5} x {f32}{\textgreater}} \\
%Convolution          & \texttt{convolve \%a:\ {\textless}\textrm{\ldots}{\textgreater}\ kernel \%b:\ {\textless}\textrm{\ldots}{\textgreater}\ stride \%c:\ {\textless}\textrm{\ldots}{\textgreater}} \\
Slice                & \texttt{slice \%a:\ {\textless}{10} x {20} x {i32}{\textgreater} from 1 upto 5} \\
%Random               & \texttt{random 768 x 10 from 0.0:\ f32 upto 1.0:\ f32} \\
%Select               & \texttt{select \%x:\ {\textless}{8} x {f64}{\textgreater},\ \%y:\ {\textless}{8} x {f64}{\textgreater} by \%z:\ {\textless}{8} x {bool}{\textgreater}} \\
Compare              & \texttt{gt \%a:\ {\textless}{10} x {20} x {bool}{\textgreater},\ \%b:\ {\textless}{1} x {20} x {bool}{\textgreater}} \\
Data type cast       & \texttt{dataTypeCast \%x:\ {\textless}{10} x {f32}{\textgreater} to f64} \\
Function application & \texttt{apply \%foo(\%x:\ f32, \%y:\ f32):\ (f32,\ f32)\ -\textgreater\ f32} \\
Branch               & \texttt{branch {\textquotesingle}block\_name(\%a:\ i32,\ \%b:\ i32)} \\
Conditional branch   & \texttt{conditional\ \%cond:\ bool\ then {\textquotesingle}bb0()\ else\ {\textquotesingle}bb1()} \\
Shape cast           & \texttt{shapeCast \%a:\ {\textless}{1} x {40} x {f32}{\textgreater}\ to\ 2\ x\ 20} \\
%Extract              & \texttt{extract \#x from \%pt:\ \$Point} \\
%Insert               & \texttt{insert\ 10:\ f32\ to\ \%pt:\ \$Point\ at\ \#x} \\
\hline
\end{tabular}

\ \\

\caption{This table illustrates a selection of the instructions in the DLVM virtual instruction set. The instruction set includes linear algebra operations such as \texttt{tanh} and \texttt{dot} in addition to control flow instructions such as \texttt{branch}.\label{tab:dlvmInstructionSet}}
\end{center}
\end{table}

%                                                                              %
%%%%%%%%%%%%%%%%%%%%%%%%%%%%%%%%%%%%%%%%%%%%%%%%%%%%%%%%%%%%%%%%%%%%%%%%%%%%%%%%

\subsection{DLVM Core \label{sec:dlvmCore}}

DLVM Core contains essential components for an optimizing compiler: 
IR, pass manager, and passes (see Figure~\vref{fig:dlvmStack}). 
The DLVM IR consists of a virtual instruction set, control flow graph and data flow representation.
Passes are functions that traverse the intermediate representation of a program, either producing useful results as analyses of the program (analysis passes), or mutating the program for differentiation and optimizations (transform passes).

\subsubsection{A Domain-specific Compiler Intermediate Representation for DLVM}

Inspired by the LLVM IR \citep{llvm} and the Swift Intermediate Language \citep{sil}, DLVM IR is a graph-based, modular code representation, with both an in-memory format and a textual format.
The code representation has a hierarchy of abstractions: \textbf{module}, \textbf{function}, \textbf{basic block}, and \textbf{instruction}.
An instruction is the minimal unit of code that operates on values, which can be globals, function arguments or temporary virtual registers produced by instructions.
Each module contains a collection of type definitions, global values and functions.
Each function has a control flow graph formed by basic blocks and control flow edges.
Each basic block contains an ordered list of instructions with data dependencies forming a directed acyclic graph.
%
%, which we call data flow graph (DFG).

The DLVM IR has a high-level type system with tensor as a first-class type.
The DLVM virtual instruction set includes domain-specific primitive math operators, as well as general-purpose instructions for memory management, control flow and function application.
Domain-specific instructions include element-wise unary operators, such as \texttt{tanh} and \texttt{negate}, element-wise binary operators, such as \texttt{add} and \texttt{power}, and complex operators such as \texttt{dot}, \texttt{transpose}, and \texttt{convolve}. All element-wise binary operators support broadcasting.
A sample of DLVM IR code is shown in Figure~\vref{fig:dlvmExampleIR}.

The DLVM instruction set does not include composite math functions such as \texttt{softmax}, \texttt{sigmoid}, \texttt{min} or \texttt{max}.
All of these functions can be composed of primitive math instructions and control flow constructs.
This design allows for the standard AD algorithm to be applied to any differentiable program, with no need for special handling of composite cases.

\subsubsection{Domain-specific Compiler Passes for DLVM}

DLVM has a full-fledged pass infrastructure, performing various analyses and two kinds of transformations: differentiation and optimization. 
Differentiation constructs function definitions from gradient declarations using adjoint code generation (see Section~\ref{sec:gradientDeclarations} below).
%
%Canonicalization lowers certain high-level instructions to lower representations that enable more optimizations. 
%
Optimization is then performed on the resulting IR, maximizing the code performance.
Optimizations include domain-specific optimizations, such as algebra simplification, linear algebra fusion, matrix multiplication reordering, and AD checkpointing, and traditional compiler optimizations.

Since DLVM IR is aware of mathematical operators such as \texttt{tanh} and \texttt{power}, the algebra simplification pass can find and simplify certain mathematical operations that are expensive or redundant.
For example, $x^2$ can be simplified to $x \odot x$ ($\odot$ stands for element-wise multiplication), and $x^0$ can be simplified to constant $1$.
Matrix multiplication reordering is another classic optimization that minimizes the number of sub-operations in a chain of matrix multiplications with different dimensionality, based on matrix multiplication’s associativity.

Since the DLVM optimizer is aware of linear algebra operations with static
dimensionality, maximizing the performance by fusing verbose linear operations
into a single matrix multiplication is beneficial as well.
For example, it is very common to encounter expressions of the form $\textbf{W}\textbf{x} + \textbf{b}$.
When unoptimized, the matrix multiplication and the addition will be parallelized separately.
Since launching compute kernels separately can be expensive, DLVM performs linear algebra fusion, which transforms subexpressions involving both matrix multiplication and element-wise operations into a single matrix multiplication instruction on padded tensors.
Besides the simple pattern like an addition of matrix multiplication and a vector, we can apply the same approach to a polynomial of multiple matrix multiplications, turning the polynomial into a single matrix multiplication.
For example, in a simple recurrent neural network (RNN), each cell of the recurrence is a feed forward neural network that takes two inputs: $\mathbf{x}_t$, the input local to the current timestep, and $\mathbf{h}_t$, the hidden state carried along the recurrence.
The linear algebra fusion pass can simplify operations in $\textbf{h}_t =
f(\textbf{W}\textbf{x}_{t-1} + \textbf{U}\textbf{h}_{t-1} + \textbf{b})$ from
two matrix multiplications and two additions into a single matrix multiplication.
A more aggressive, interprocedural version of linear algebra fusion can optimize parameter passing and memory allocation, so that the entire concatenated matrix can be created and passed around in the first place without reallocation.

%%%%%%%%%%%%%%%%%%%%%%%%%%%%%%%%%%%%%%%%%%%%%%%%%%%%%%%%%%%%%%%%%%%%%%%%%%%%%%%%
%                                                                              %
%      Figure: Sample MNIST program in DLVM intermediate representation        %
%                                                                              %
                              \begin{figure}
\begin{center}
\begin{Verbatim}[frame=single,fontsize=\small,commandchars=\\\{\}]
\textcolor{xcodeKeyword}{module} "my_module"
\textcolor{xcodeKeyword}{stage} raw

\textcolor{xcodeComment}{// Representing function foo(x, w, b) = dot(x, w) + b}
\textcolor{xcodeKeyword}{func} @foo: (<1 x 784 x \textcolor{xcodeType}{f32}>, <784 x 10 x \textcolor{xcodeType}{f32}>, <1 x 10 x \textcolor{xcodeType}{f32}>)
          -> <1 x 10 x \textcolor{xcodeType}{f32}> {\{}           
{\textquotesingle}entry(%x: <1 x 784 x \textcolor{xcodeType}{f32}>, %w: <784 x 10 x \textcolor{xcodeType}{f32}>, %b: <1 x 10 x \textcolor{xcodeType}{f32}>):
    %v0 = dot %x: <1 x 784 x \textcolor{xcodeType}{f32}>, %w: <784 x 10 x \textcolor{xcodeType}{f32}>
    %v1 = add %v0: <1 x 10 x \textcolor{xcodeType}{f32}>, %b: <1 x 10 x \textcolor{xcodeType}{f32}>
    \textcolor{xcodeKeyword}{return} %v1: <1 x 10 x \textcolor{xcodeType}{f32}>
{\}}

\textcolor{xcodeComment}{// Gradient of @foo with respect to all arguments}
[\textcolor{xcodeKeyword}{gradient} @foo]
\textcolor{xcodeKeyword}{func} @foo_grad: (<1 x 784 x \textcolor{xcodeType}{f32}>, <784 x 10 x \textcolor{xcodeType}{f32}>, <1 x 10 x \textcolor{xcodeType}{f32}>) 
           -> (<1 x 784 x \textcolor{xcodeType}{f32}>, <784 x 10 x \textcolor{xcodeType}{f32}>, <1 x 10 x \textcolor{xcodeType}{f32}>)

\textcolor{xcodeComment}{// Gradient of @foo with respect to arguments 1 and 2}
\textcolor{xcodeComment}{// Keeping original output 0}
\textcolor{xcodeComment}{// Seedable, able to take back-propagated gradient as a seed for AD}
[\textcolor{xcodeKeyword}{gradient} @foo \textcolor{xcodeKeyword}{wrt} 1, 2 \textcolor{xcodeKeyword}{keeping} 0 \textcolor{xcodeKeyword}{seedable}]
\textcolor{xcodeKeyword}{func} @foo_grad_3:
    (<1 x 784 x \textcolor{xcodeType}{f32}>, <784 x 10 x \textcolor{xcodeType}{f32}>, <1 x 10 x \textcolor{xcodeType}{f32}>, <1 x 10 x \textcolor{xcodeType}{f32}>)
   -> (<784 x 10 x \textcolor{xcodeType}{f32}>, <1 x 10 x \textcolor{xcodeType}{f32}>, <1 x 10 x \textcolor{xcodeType}{f32}>)
\end{Verbatim}

\end{center}
\caption{Example code in DLVM intermediate representation. Note that some functions are annotated as defining the \texttt{gradient} of another function with respect some or all arguments. The body of these gradient functions will be automatically generated. \label{fig:dlvmExampleIR}}
\end{figure}

%\textcolor{xcodeComment}{// Gradient of @foo with respect to arguments 1 and 2}
%[\textcolor{xcodeKeyword}{gradient} @foo \textcolor{xcodeKeyword}{wrt} 1, 2]
%\textcolor{xcodeKeyword}{func} @foo_grad_2: (<1 x 784 x \textcolor{xcodeType}{f32}>, <784 x 10 x \textcolor{xcodeType}{f32}>, <1 x 10 x \textcolor{xcodeType}{f32}>)
%                 -> (<784 x 10 x \textcolor{xcodeType}{f32}>, <1 x 10 x \textcolor{xcodeType}{f32}>)

%                                                                              %
%%%%%%%%%%%%%%%%%%%%%%%%%%%%%%%%%%%%%%%%%%%%%%%%%%%%%%%%%%%%%%%%%%%%%%%%%%%%%%%%

\subsubsection{Algorithmic Differentiation through Adjoint Code Generation \label{sec:gradientDeclarations}}

Algorithmic differentiation (AD), also known as automatic differentiation, encompasses a family of a well-known techniques for algorithmically obtaining the derivatives of a function $f : \mathbf{x} \in \mathbb{R}^n \rightarrow \mathbf{y} \in \mathbb{R}^m$ \citep{AD}.
The function $f$ can be represented as a directed acyclic computation graph representing the composition of elementary computational operations for which the respective derivatives are well known.
The partial derivative $\frac{\partial y_j}{\partial x_i}$ can be computed through recursive applications of the chain rule, either in the forward direction (corresponding to a bottom-up traversal of the computation graph) or in the backward direction (corresponding to a top-down traversal of the computation graph).
The former approach, called forward-mode or tangent-mode AD, is used in some research communities, most commonly when $m \gg n$ \citep{2016:Goodfellowetal}.
The latter approach, which includes the back-propagation algorithm \citep{1986:RumelhartMcClelland} as a special case, is called reverse-mode or adjoint-mode AD, and encompasses the techniques most commonly used for training the weights in neural networks.

In DLVM, the differentiation pass is responsible for performing reverse-mode AD.
This pass is responsible for generating DLVM IR code that calculates the derivative of a differentiable function.
%
%The first and most important compiler pass is reverse-mode algorithmic differentiation, implemented in the Differentiation pass.
%
A function is marked as being automatically differentiable via \textbf{gradient declarations}.
A gradient declaration is a function in a module that is declared with its mathematical relation with another function in the module and no function body.
The function \texttt{@foo\_grad} in Figure~\ref{fig:dlvmExampleIR} is an example of such a function. Gradient declarations are configurable, e.g. specifying arguments to differentiate with respect to, keeping original outputs, and toggling seedability to accept back-propagated gradients.
The differentiation pass, when applied, canonicalizes every gradient declaration in the module to a normal function definition with basic blocks and instructions. The canonicalization process first copies basic blocks and instructions from the original function to the new function body, and then applies adjoint code generation to the function.
Unlike many of the existing deep learning frameworks, AD in DLVM is a source code transformation, not interpretation (operator overloading) over the same program.
This makes the compiler able to perform optimizations on the gradient computation separately and enables higher order differentiation. %, a good example of the separation of concerns.

Given a differentiable function $f(\textbf{x}_1, \textbf{x}_2, \ldots, \textbf{x}_n)$, this pass creates a new function that computes the Jacobian $\textbf{J}_{f}$.
This approach to AD has several advantages with respect to AD performed by operator overloading / graph interpretation.
Unlike operator overloading, the gradient function produced by AD is a standalone function that sits uniformly alongside other functions in an IR module, representationally unrelated to the original function.
The generated function takes original inputs and produces a tuple of partial derivatives with respect to the inputs.
In AD, not all values in the forward evaluation will necessarily be used to compute derivatives.
In DLVM, unused operations can be easily eliminated by the aggressive dead code elimination pass in the compilation pipeline (see Section \ref{sec:generalPurposeOpt}).
In addition, an AD-specific optimization technique called checkpointing can further reduce memory consumption during gradient computation.

AD in DLVM is configurable.
The front-end can choose to differentiate a function with respect to selected arguments, to keep some of the outputs of the original function, to apply differentiation to a specific output when there are multiple return values, or to enable the function to accept back-propagated gradients (seeds) through function composition, all by gradient declarations. If the function returns multiple values in a tuple, the gradient declaration can also specify which tuple element to differentiate.
Our approach to AD is implemented as a transformation from one function to another function.
This approach also makes higher-order differentiation possible;
this can be accomplished by declaring a higher-order gradient function that differentiates the original gradient function.
%
%In deep learning, it is sometimes desirable to take the second-order gradient of a loss function with respect to neural network parameters in order to observe the rate of gradient descent, eliminate superfluous weights, and estimate confidence intervals for both weights and network outputs \citep{1994:Buntine}.

\subsubsection{General-purpose Optimizations for DLVM \label{sec:generalPurposeOpt}}

General-purpose optimizations refer to traditional compiler optimizations applied to DLVM IR.
These optimizations are important at the DLVM stage in the compilation pipeline, since linear algebra computation can be highly optimized or eliminated before they get lowered to LLVM IR which contain parallel execution and low-level information that prevent LLVM optimizations from identifying high-level patterns.
Some of these optimizations are aggressive dead code elimination, common subexpression elimination, and sparse conditional constant propagation.
Applying such optimizations on gradient computation is not feasible in other approaches to AD that use graph interpretation (operator overloading), because the forward pass and the backward pass are tied together in a single graph; mutating either evaluation pass will alter the semantics of the other.

\subsection{Code generation \label{sec:dlvmCodeGen}}

Two major design goals of DLVM are the ability to target multiple heterogenous parallel architectures from the same front-end DSL code (and corresponding DLVM IR), and the ability to perform aggressive optimizations on lowered programs.
In order to attain these goals, DLVM code generation transforms DLVM IR into LLVM IR.
LLVM is a robust and mature compiler infrastructure with multiple back-ends, including NVIDIA GPUs.
Many high-level DLVM IR linear algebra instructions over tensors abstract lower-level operations. 
The DLVM compiler transforms the high-level DLVM IR into lower-level stages and ultimately into calls to BLAS and compute kernels in LLVM IR.
Existing LLVM utilities are used to compile the generated LLVM IR to the final binary.

In order to take full advantage of a variety of emerging heterogeneous parallel architectures, we plan for future versions of DLVM to target the IR of HPVM \citep{hpvm}, a higher-level heterogeneous compiler extension to LLVM IR that allows for transparent targeting of diverse architectures from a data flow graph.

\subsection{DLVM command line toolchain \label{sec:dlvmCommandLine}}

The front-end software associated with each DSL (see Section~\ref{sec:DSLs}) is responsible for generating a DLVM IR for a given source language program to be compiled.
The DLVM compiler infrastructure itself is a compiler from DLVM IR to LLVM IR, therefore having a command line toolchain is necessary for verifying, transforming and compiling batches of DLVM IR files (\texttt{*.dl}).
Unlike XLA and NNVM/TVM, which only provide a Python/C++ interface to their users, DLVM provides a command line interface like any industry-standard compiler.

The DLVM optimizer utility, \texttt{dlopt}, accepts \texttt{*.dl} IR files and applies user-specified optimization passes on them. 
The DLVM compiler driver, \texttt{dlc}, accepts \texttt{*.dl} IR files and performs user-specified tasks, such as verification, differentiation, optimization passes, stage lowering passes, and code generation; the driver invokes the DLVM core library to achieve these tasks.
Because of having a textual format of the IR, the DLVM framework can easily make use of the LLVM Integrated Tester (lit) and FileCheck to perform robust unit testing. In future development, we plan to introduce a DLVM bitcode format for compact storage and high-throughput processing of DLVM code.

%%%%%%%%%%%%%%%%%%%%%%%%%%%%%%%%%%%%%%%%%%%%%%%%%%%%%%%%%%%%%%%%%%%%%%%%%%%%%%%%
%                                                                              %
%                             Example code in NNKit                            %
%                                                                              %
                              \begin{figure}
\begin{center}
\begin{Verbatim}[frame=single,fontsize=\small,commandchars=\\\{\}]
\textcolor{xcodeComment}{// Staged function representing f(x, w, b) = dot(x, w) + b}
\textcolor{xcodeKeyword}{let} f: \textcolor{xcodeType}{Rep}<(\textcolor{xcodeType}{Float2D}, \textcolor{xcodeType}{Float2D}, \textcolor{xcodeType}{Float2D}) -> \textcolor{xcodeType}{Float2D}> = 
    lambda \textcolor{black}{\{} x, w, b \textcolor{xcodeKeyword}{in}
            x • w + b
    \textcolor{black}{\}}

\textcolor{xcodeComment}{// Staged function 'g', type-inferred from 'f'}
\textcolor{xcodeKeyword}{let} g = lambda \textcolor{black}{\{} x, w, b in
    \textcolor{xcodeKeyword}{let} linear = f[x, w, b] \textcolor{xcodeComment}{// staged function application}
    \textcolor{xcodeKeyword}{return} tanh(linear)
\textcolor{black}{\}}

\textcolor{xcodeComment}{// Gradient of 'g' with respect to arguments 'w' and 'b'}
\textcolor{xcodeKeyword}{let} dg = gradient(of: g, withRespectTo: (1, 2), keeping: 0)
\textcolor{xcodeComment}{// 'dg' has type:}
\textcolor{xcodeComment}{// Rep<(Float2D, Float2D, Float2D) -> (Float2D, Float2D, Float2D)>}

\textcolor{xcodeComment}{// Call staged function on input data 'x', 'w' and 'b'}
\textcolor{xcodeKeyword}{let} (dg_dw, dg_db, result) = dg[x, w, b]
\textcolor{xcodeComment}{// At runtime, 'dg' gets just-in-time compiled though DLVM,}
\textcolor{xcodeComment}{// and computes ( dg/dw, dg/db, g(x, w, b) )}

\textcolor{xcodeComment}{// Second order derivative of 'g' with respect to 'w'}
let d2g_dw2 = gradient(of: dg, from: 0, withRespectTo: (1))
\textcolor{xcodeComment}{// 'd2g_dw2' has type:}
\textcolor{xcodeComment}{// Rep<(Float2D, Float2D, Float2D) -> Float2D>}
\end{Verbatim}

\end{center}
\caption{Example code in Swift using NNKit, a staged DSL targeting DLVM. \label{fig:dlvmExampleNNKit}}
\end{figure}

%                                                                              %
%%%%%%%%%%%%%%%%%%%%%%%%%%%%%%%%%%%%%%%%%%%%%%%%%%%%%%%%%%%%%%%%%%%%%%%%%%%%%%%%

\subsection{Neural network DSLs \label{sec:DSLs}}

Most of existing deep learning DSLs are embedded in a dynamically typed scripting language such as Python and Lua. While these scripting languages provide flexibility and a large number of libraries for scientific computing, they can act as a barrier between lightweight prototyping code and systematic production code. This barrier significantly reduces the reliability of ML software, resulting in suboptimal programming experience and unnecessarily effortful development cycles.

In software engineering, a proven approach to tackle this problem is language and compiler technologies, starting from a language that is amenable to static analysis. A well-designed deep learning DSL should support the needs of deep learning software developers by providing a safe environment for rapid prototyping, while simultaneously generating highly efficient code for training and inference. DSLs in a scripting language can easily achieve rapid prototyping, but they are generally incapable of providing a safe environment with optimal performance. We believe that the best solution is DSLs embedded in a type-safe, type-inferring programming language that is both fast and easy to learn. In our initial release of DLVM, we provide one such DSL, both as a proof-of-concept and as a reference implementation that showcases the capabilities of DLVM as a platform for deep learning DSL development.

NNKit is a staged DSL embedded in Swift, featuring natural expression of tensor computation alongside the host program without losing static analyses and optimizations on the tensor program. Inspired by Lightweight Modular Staging \citep{lms}, NNKit leverages the static type system of Swift to guarantee type safety of the DSL. Tensor types are wrapped in \texttt{Rep<T>}, meaning the representation of some computation that produces data of type \texttt{T}. Tensor operators overloaded for \texttt{Rep} are essentially AST builders for delayed evaluation. Instead of generating computation nodes at runtime and performing operator-overloading AD like PyTorch or TensorFlow Eager \citep{tfeager}, NNKit tensor computations are staged once during the lifetime of the host program. At invocation time of staged functions, NNKit emits shape-specialized DLVM IR and leverages DLVM to perform AD, optimizations, and low-level code generation. A sample of Swift code using NNKit is shown in Figure~\vref{fig:dlvmExampleNNKit}.

The NNKit just-in-time compiler has four important phases:
The first phase, expression staging, produces an unshaped graph IR of the tensor computation.
The second phase, shape specialization, prepares to generate statically shaped DLVM IR for staged functions when they are applied to shaped tensors.
The third phase, lowering, generates DLVM IR and passes it through DLVM, producing a dynamic library containing a function symbol.
The final phase, function reification, loads the binary and wraps the low-level function to a Swift function.

We anticipate other existing deep learning frameworks, such as TensorFlow, could be adapted to use DLVM as a back-end to their tensor math DSLs.

\section{Conclusion}

The deep learning research community has a rich variety of available frameworks.
While two existing projects have attempted a compilers approach to deep learning frameworks, and have respectively achieved good integration with existing systems (TensorFlow XLA) and good performance (NNVM + TVM), their design philosophies have not entirely followed established best practices in optimizing compiler design.
While well intentioned, the remaining vast majority of other frameworks have failed to observe that the problem of front-end DSLs, algorithmic differentiation, and converting a neural network into efficient executable code is, at its core, a compilers problem. 
As a result, important issues of extensibility and optimization have been addressed in less than optimal fashion in such frameworks.
Nevertheless, several such frameworks have achieved wide adoption. 
We believe that the principled application of optimizing compiler techniques will lead to substantial improvements in the tools available to deep learning researchers.
DLVM and its associated front-end DSLs have a major role to play in this future.
Our existing implementation supports reverse-mode AD in the core language, and utilizes LLVM to target NVIDIA GPUs.
In our ongoing work, we plan to substantially increase the number of supported hardware architectures by utilizing HPVM as an additional back-end, and explore more advanced AD techniques such as mixing forward and reverse modes.

\bibliography{main}
\bibliographystyle{iclr2018_workshop}

\end{document}